\documentclass[prb,twocolumn,amsmath,amssymb,superscriptaddress]{revtex4-1}
\usepackage{graphicx}% Include figure files
\usepackage{dcolumn}% Align table columns on decimal point
\usepackage{hhline}
\usepackage{bm,color}% bold math

\usepackage{float}

\usepackage{amsmath}

\usepackage{epstopdf}

\begin{document}

%\wideabs{

\title{Putative spin liquid in the triangle-based iridate Ba$_3$IrTi$_2$O$_9$}

\author{W.-J. Lee}
\affiliation{Department of Physics, Chung-Ang University,  Seoul 156-756, Republic of Korea}

\author{S.-H. Do}
\affiliation{Department of Physics, Chung-Ang University,  Seoul 156-756, Republic of Korea}

\author{Sungwon Yoon}
\affiliation{Department of Physics, The Catholic University of Korea, Bucheon 420-743, Republic of Korea}

\author{S. Lee}
\affiliation{Department of Physics, Chung-Ang University,  Seoul 156-756, Republic of Korea}

\author{Y.S. Choi}
\affiliation{Department of Physics, Chung-Ang University,  Seoul 156-756, Republic of Korea}

\author{D.J. Jang}
\affiliation{Max Planck Institute for Chemical Physics of Solids
Dresden, Germany}

\author{M. Brando}
\affiliation{Max Planck Institute for Chemical Physics of Solids
Dresden, Germany}

\author{M. Lee}
\affiliation{National High Magnetic Field Laboratory, Florida State University, Tallahassee, FL 32310, USA}

\author{E. S. Choi}
\affiliation{National High Magnetic Field Laboratory, Florida State University, Tallahassee, FL 32310, USA}

\author{S. Ji}
\affiliation{Max Planck POSTECH Center for Complex Phase Materials, Pohang 37673, Republic of Korea}

\author{Z. H. Jang}
\affiliation{Department of Physics, Kookmin University, Seoul 136-702, Republic of Korea}

\author{B. J. Suh}
\affiliation{Department of Physics, The Catholic University of Korea, Bucheon 420-743, Republic of Korea}

\author{K.-Y. Choi}
\email[]{kchoi@cau.ac.kr}
\affiliation{Department of Physics, Chung-Ang University,  Seoul 156-756, Republic of Korea}

%\date{\today}

\begin{abstract}
We report on thermodynamic, magnetization, and muon spin relaxation measurements of
the strong spin-orbit coupled iridate Ba$_3$IrTi$_2$O$_9$, which constitutes a new frustration motif made up a mixture of edge- and corner-sharing triangles. In spite of strong antiferromagnetic exchange interaction of the order of 100~K,
we find no hint for long-range magnetic order down to 23 mK.
The magnetic specific heat data unveil the $T$-linear and -squared dependences at low temperatures below 1~K.
At the respective temperatures, the zero-field muon spin relaxation features a persistent spin dynamics, indicative of unconventional low-energy excitations. A comparison to the $4d$ isostructural compound Ba$_3$RuTi$_2$O$_9$  suggests that a concerted interplay of compass-like magnetic interactions and frustrated geometry  promotes a dynamically fluctuating state in a triangle-based iridate.
\end{abstract}

\maketitle

\section{Introduction}
Quantum spin liquids (QSLs) are a quantum paramagnetic state
that entails fractionalized quasiparticles, quantum entanglement, and topological order~\cite{Balents,Savary}.
This novel state of matter is characterized by
a macroscopic degeneracy of ground states that triggers fluctuations
among several possible spin configurations, thereby preventing a symmetry breaking even
at zero temperature.  In search of QSLs, the local connectivity of edge- and corner-sharing triangles offers an elementary motif of geometrically frustrated magnets. The best physical realizations of QSLs known to date are the organic triangular compounds  $\kappa$-(BEDT-TTF)$_2$Cu$_2$(CN)$_3$~\cite{Shimizu,Yamashita08,Yamashita09,Pratt} and EtMe$_3$Sb[Pd(dmit)$_2$]$_2$~\cite{Yamashita10,SYamashita}, the rare-earth triangular system
YbMgGaO$_4$~\cite{Shen,Paddison}, and the kagome antiferromagnet ZnCu$_3$(OH)$_6$Cl$_2$~\cite{Mendels,Han,Fu},
which commonly show the absence of static magnetism and the presence of fractionalized spinon (charge neutral $S=1/2$) excitations.

On the other hand, strong spin-orbit coupled magnetic insulators
offer another efficient route to achieve a QSL ground state. Owing to exchange frustration, a bond-dependent Kitaev honeycomb lattice
harbors a topological QSL with emergent gapless and gapped Majorana-fermion excitations~\cite{Kitaev,Baskaran,Knolle}.
The promising candidates for Kitaev physics are the harmonic-honeycomb iridates $\alpha$-, $\beta$-, and $\gamma$-Li$_2$IrO$_3$~\cite{Singh,Takayama,Modic,Glamazda} and the ruthenium tricloride $\alpha$-RuCl$_3$~\cite{Plumb,Sandilands,Banerjee}.  An ensuing question to ask is whether an exotic state of matter emerges
when two central threads of frustration are combined into a single system.

The 6H-perovskite compounds Ba$_3$MA$_2$O$_9$ (M=transition metals; A=Sb, Ti)
are an outstanding class of materials to explore the synergy effect of geometrical and exchange frustration thanks to their alternating arrangement of magnetic and nonmagnetic triangular lattices as well as to their capacity of hosting
a vast library of chemical variants, ranging from  a $3d$ to $5d$ metal ion~\cite{Zhou,Nakatsuji,Corboz,Susuki,Cheng,Dey,Dey13}.
Depending on the metal cation M, diverse ground states have been reported, including a spin-orbit liquid state, spin-freezing, 120$^{\circ}$ order, and a putative spin liquid. Very recently, Ba$_3$IrTi$_2$O$_9$ has received  much theoretical interest in an attempt to
extend a Kitaev physics to a case of triangular iridates~\cite{Kimchi,Becker,Catuneanu}.
Based on {\it ab initio}  calculations, A. Catuneanu {\it et al.}~\cite{Catuneanu} evaluated a minimal spin Hamiltonian
with the Kitaev interaction $K\simeq -4$~meV, the Heisenberg interaction $J\simeq -2$~meV,
and the symmetric off-diagonal interaction $\Gamma\simeq 2$~meV.
However, the Ir/Ti site disorders prevent the realization of the  $J-K-\Gamma$ triangular lattice model in Ba$_3$IrTi$_2$O$_9$.
Nevertheless, as exemplified in Ba$_3$CuSb$_2$O$_9$~\cite{Nakatsuji}, a new structural motif could emerge from the original triangular lattice thanks to the apparent structural imperfection, thereby stabilizing a novel state of matter~\cite{Furukawa}.

In this paper, we introduce the spin-orbit coupled Ba$_3$MTi$_2$O$_9$ (M=Ru,Ir) as a structural variant of
the  $J-K-\Gamma$ triangular lattice. Using specific heat, magnetization, and muon spin resonance, we show experimental signatures
of a spin liquid in M=Ir, namely, coherently fluctuating spins down to 23~mK
and spinon-like unconventional excitations.

\section{\label{sec:level2}Experimental Details}
The polycrystalline samples of Ba$_3$MTi$_2$O$_9$~(M=Ru, Ir) were synthesized by a solid state reaction method. Stoichiometric amounts of BaCO$_3$, RuO$_2$, IrO$_2$, and TiO$_2$ were thoroughly ground and mixed, and the resulting mixtures were made into pellets. The pellets were sintered at 1100~$^{\circ}$C for 72 hours with intermediate grindings.  Powder x-ray diffraction measurements were carried out at room temperature by PANalytical Empyrean ALPHA-1 diffractometer
with graphite monochrometer (Cu $K_\alpha$; $\lambda=1.54056$ {\AA}) for the Ir sample and Bruker D8 Advance diffractometer
with Cu $K_\alpha$ radiation ($\lambda=1.54182$ {\AA}) for the Ru sample. $dc$ magnetization was measured in the temperature range $T=2-300$~K using conventional SQUID magnetometer (MPMS, Quantum Design). Electric resistivity of the Ba$_3$MTi$_2$O$_9$ pellets was measured using DC resistivity option in a Quantum Design physical property measurement system (PPMS).

Specific heat measurements were carried out using a semi-adiabatic compensated heat-pulse method described in Ref. ~\cite{Wilhelm}. Because of the low heat capacity of the sample, particular attention has been paid in the measurement and subtraction of the addenda which consist of the heat capacity of the silver platform, thermometer, heater, connecting wires and a very small amount of grease used to attach the sample. The error bars were calculated from the error of the semiadiabatic heat-pulse method itself and the temperature deviation of the platform thermometer from a calibrated reference thermometer.
Muon spin relaxation ($\mu$SR) experiments were performed on the DR spectrometer ($T=0.02-4$~K) and the LAMPF spectrometer ($T=1.6 - 125$~K) at TRIUMF (Vancouver, Canada).  The collected $\mu$SR data were analyzed by using the free
software package MUSRFIT~\cite{Suter}.

\section{Results and discussion}
\subsection{Spin topology and basic properties}

%%%%%%%%%%%%%%%%%%%%%%%%%%%%%%%%%%%%%%%%%%%%%%%%%%%%%%%%%%%%%%%%%%%%%%%%%%%%%%%%%%
%%%%%%%%%%%%%%%%%%%%%%%%%%%%%%%%%%%%%%%%%%%%%%%%%%%%%%%%%%%%%%%%%%%%%%%%%%%%%%%%%

\begin{figure}
\label{figure1}
\centering
\includegraphics[width=8cm]{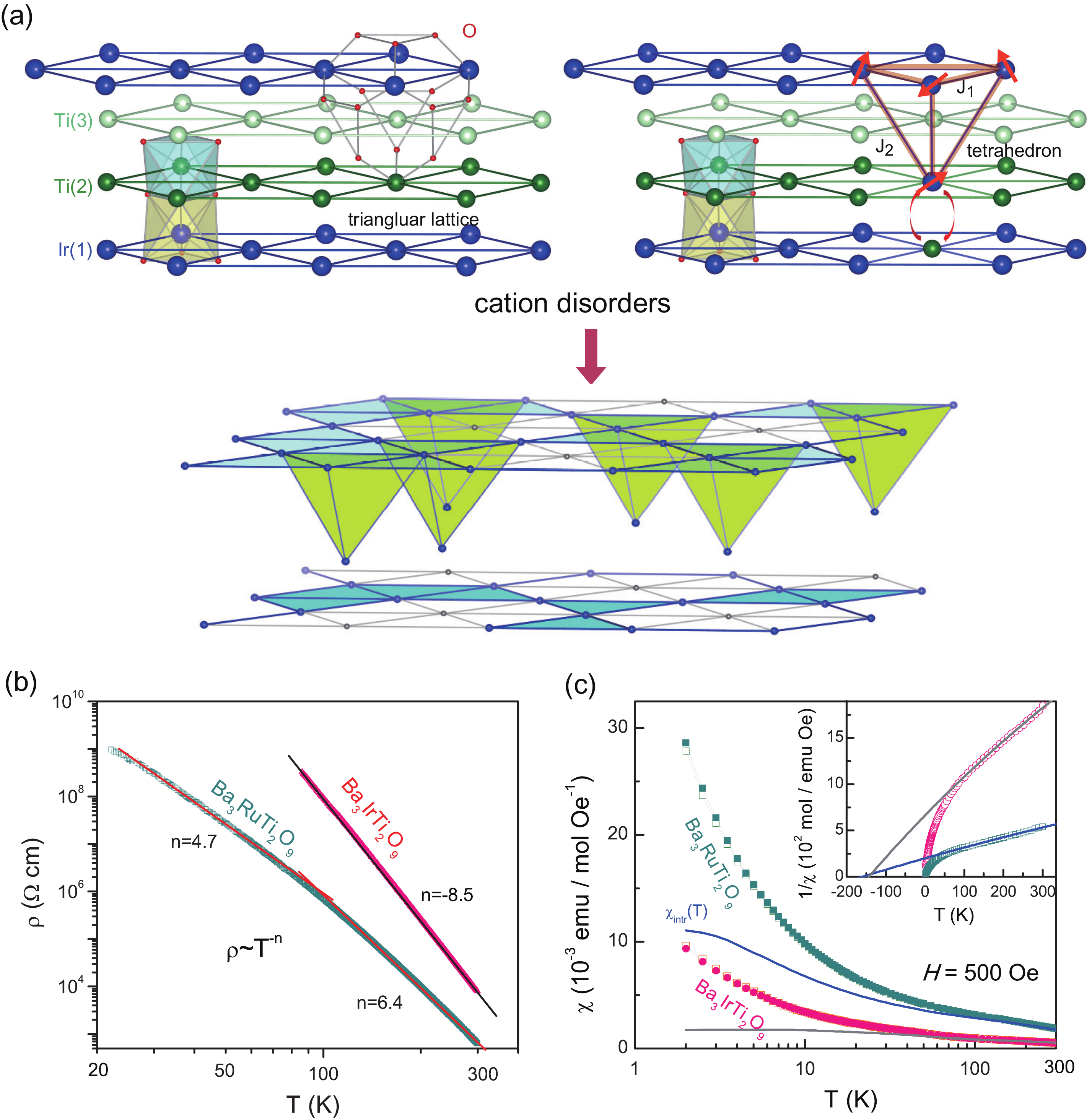}
\caption{
(a) Triangular bilayers of Ir(1)$^{4+}$/Ti(2)$^{4+}$ which are separated by a nonmagnetic triangular layer of Ti(3)$^{4+}$ and
 are stacked along the $c$-axis. The blue, green, and red balls stand for Ir, Ti, and O atoms, respectively.  The barium atoms are omitted for clarity. The gray solid lines depict the superexchange paths Ir-O-O-Ir.
In the presence of the Ti$^{4+}$/Ir$^{4+}$ cation disorders within the Ir(1)Ti(2)O$_9$ dimers, a nearly isotropic tetrahedron ($J_1\approx J_2$) is created.  (b) Temperature dependence of resistivity of Ba$_3$MTi$_2$O$_9$ plotted on a log-log scale. The solid lines are fits
to a power law $\rho(T)\sim T^{-n}$. (c) Temperature dependence of the magnetic susceptibility of Ba$_3$MTi$_2$O$_9$ measured at $H=500$~Oe in zero-field-cooled and
field-cooled processes. The solid lines represent the intrinsic magnetic susceptibility $\chi_{\mathrm{intr}}(T)$ obtained by subtracting the contribution from orphan spins. The inset plots the inverse magnetic susceptibility. The Curie-Weiss fits  are shown in the temperature range $T=100 - 320$~K with solid lines.}
\end{figure}

Figure~1(a) presents the elementary building block of Ba$_3$IrTi$_2$O$_9$, that is, the structural Ir(1)Ti(2)O$_9$ dimers
made of face-sharing of Ir(1)O$_6$ and Ti(2)O$_6$ octahedra. These pairs form triangular bilayers of
Ir$^{4+}$ ($5d^5$; $J_\mathrm{eff}=1/2$) and Ti$^{4+}$ ($S=0$), stacked along the crystallographic $c$ axis and separated by another nonmagnetic triangular layer of Ti(3)$^{4+}$.  Because of the similar ionic radii between Ir$^{4+}$ and
Ti$^{4+}$ ions, the Ir/Ti intersite disorders  are unavoidable. Indeed, our x-ray diffraction measurements uncover a random occupation of about 39 \% nonmagnetic Ti$^{4+}$ ions in the magnetic triangular layers and
a few percentage of Ir$^{4+}$ ions in the single non-magnetic Ti(3) layer (see Supplementary Materials A and B~\cite{SM}).  Ba$_3$RuTi$_2$O$_9$
shares the same spin topology as Ba$_3$IrTi$_2$O$_9$~\cite{Radtke}.

At a first glance, the cation disorder seems to generate two decoupled 1/3 depleted
and 2/3 depleted triangle layers. In this conventional scenario, the 39~\% spins created in the nonmagnetic triangular subsystem would exceed the percolation threshold of $p_c=0.5$ expected for a 2D triangular lattice.
Contrary to the expected 39~\% orphan spins, the high-field magnetization data place an upper limit on the magnetic impurities of about 2~\%  (see Sec. III B). This means that despite the apparent dilution,  most of Ir$^{4+}$ spins
are strongly exchange-coupled with the exchange interaction strength larger than 100~K.

This conundrum is solved by considering the out-of-plane exchange interaction $J_2$($\approx J_1$) through Ir-O-O-Ir exchange paths, which is absent in the original triangular lattice but is activated when the M(1) atoms go into the Ti(2) site. We refer to Fig.~1(b) and Table II of Supplementary Material for
the distances and angles of the two different $J_1$ and $J_2$ exchange paths~\cite{SM}. As the neighboring triangular bilayers are shifted with respect to the other such that the
vertices of the first bilayer are at the center of the second one, the Ir/Ti cation disorders within the structural dimers supply a structural motif to form a tetrahedron magnetic lattice. Given that the M$^{4+}$ spins are distributed in the 1:2 ratio between
the M(1) and Ti(2) triangular plane, the magnetic lattice of Ba$_3$MTi$_2$O$_9$ forms the 1/4 depleted
tetrahedral lattice, viewed as a combination of corner- and edge-sharing triangles. It should be noted that a  ground state degeneracy of the 1/4 depleted
tetrahedral lattice in two dimensions is something between the triangular and the kagome lattices. Therefore, the structural imperfection allows investigation of an unexplored frustration motif that bridges the triangular and kagome lattice.

Before proceeding, we stress that the interlayer interaction $J_2$ should be always invoked in the presence of
the M/A site disorders for the Ba$_3$MA$_2$O$_9$ family. For instance, the newly activated $J_2$  interaction gives a natural
explanation why  the Curie-Weiss temperature varies hardly with the concentration of the M ions in Ba$_3$M$_x$Ti$_{3-x}$O$_9$ (M = Ir, Rh)~\cite{Kumar}.

%%%%%%%%%%%%%%%%%%%%%%%%%%%%%%%%%%%%%%%%%%%%%%%%%%%%%%%%%%%%%%%%%%%%%%%%%%%%%%%%%%
%%%%%%%%%%%%%%%%%%%%%%%%%%%%%%%%%%%%%%%%%%%%%%%%%%%%%%%%%%%%%%%%%%%%%%%%%%%%%%%%%

In Fig.~1(b), the temperature dependences of resistivity $\rho(T)$ of the Ba$_3$MTi$_2$O$_9$ pellets are plotted on a double logarithmic scale.  At room temperature the resistivity is an order of $10^{3-4}$~$\Omega$cm and increases
up to $10^{9}$~$\Omega$cm at low temperatures. Overall, $\rho(T)$ displays a power-law
divergence $\rho(T)\propto T^{-n}$, establishing that all samples are a Mott insulator.
The exponent $n=6.4$ for the Ru compound (changing to $n=4.7$ at about 90 K) increases to $8.5$ for the Ir compound, suggesting that Mottness becomes stronger with increasing spin-orbit coupling. This highly insulating behavior corroborates that localized
magnetic moments are responsible for the magnetism, excluding any role of charge fluctuations.

The magnetic behavior of the Ir$^{4+}$/Ru$^{4+}$ ($4d^4;S=1$) local moments was investigated by magnetic susceptibility $\chi(T)$. As shown in Fig.~1(c), the $T$ dependence of $\chi(T)$ displays a Curie-Weiss behavior in the temperature range $T=100-320$~K and a strong enhancement at low temperatures with no hint of long range magnetic order. Fittings to the Curie-Weiss formula yield  the effective moment $\mu_{\mathrm{eff}}=2.5\, \mbox{and}\, 1.3$~$\mu_\mathrm{B}$, and the Weiss temperature $\Theta_{\mathrm{CW}}=-160$ and $-143$~K for the Ru and Ir compound, respectively, indicative of strong antiferromagnetic interactions [see the inset of Fig.~1(c)].
The steeply increasing $\chi(T)$ at low temperatures is due to a few percentage of orphan spins. After subtracting their contribution (see Sec. III B for detailed estimate), the intrinsic magnetic susceptibility $\chi_{\mathrm{intr}}(T)$ is plotted in Fig.~1(c). On cooling towards 4~K, $\chi_{\mathrm{intr}}(T)$ tends to
saturate to a constant value.

A close inspection of $\chi(T)$ unravels a small splitting between the zero-field cooled (ZFC) and field-cooled (FC) data
at low temperatures for  both compounds. Based on the $ac$ susceptibility
of Ba$_3$RuTi$_2$O$_9$ that exhibits a peak at $T=67$~mK and its frequency dependence (see Supplementary Material C~\cite{SM}),
we conclude that the Ru compound  undergoes a transition to a spin frozen or glasslike state.
In contrast, the $ac$ susceptibility of Ba$_3$IrTi$_2$O$_9$ exhibits neither sharp features nor frequency dependence.
Obviously, this rules out the presence of slow dynamics.
However, the existence of a dynamically fluctuating state in the Ir compound cannot be validated on the sole basis of
the $dc$ and $ac$ susceptibility measurements.

\subsection{Magnetization}
\begin{figure}
\includegraphics[width=9cm]{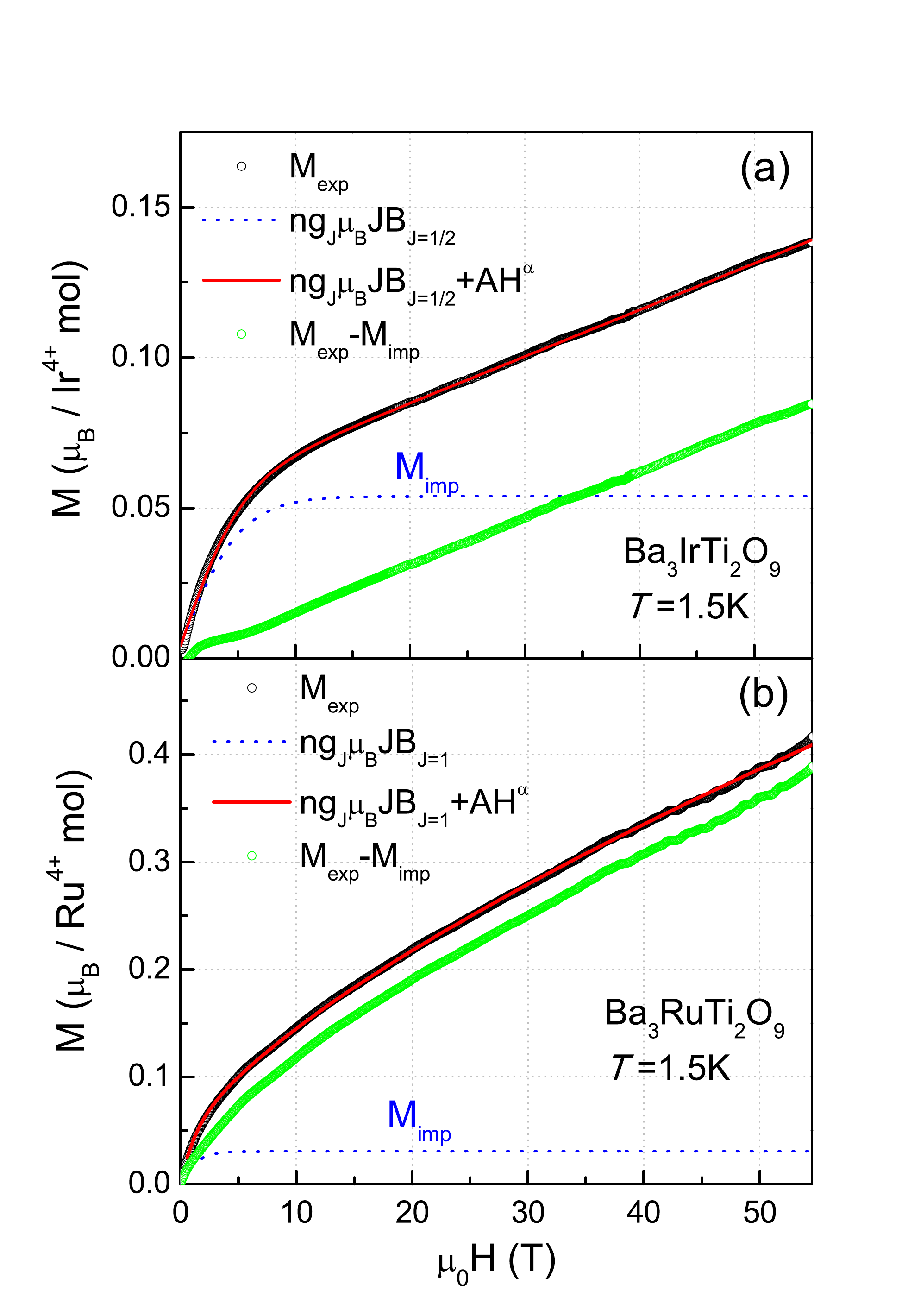}
\caption{Magnetization curve $M(H)$ of  Ba$_3$IrTi$_2$O$_9$ and Ba$_3$RuTi$_2$O$_9$ measured at $T=1.5$~K. The  gray circles are the experimental data and the red solid line is a fit of $M(H)$ to the equation described in the text. The blue dashed lines are the contribution of orphan spins to magnetization, $M_\textrm{imp}(H)$, and the green open circles represent the residual magnetization intrinsic to the system obtained by subtracting the orphan spin contribution
from the raw data.\label{fig3}}
\end{figure}

In order to differentiate disorder induced defects from the intrinsic magnetic susceptibility $\chi_{\mathrm{intr}}(T)$,
we performed high-field magnetization measurements at Dresden High Magnetic Field Laboratory using a pulsed
field magnet with 20-ms duration and an induction method with a pick-up coil device. Shown in Fig.~\ref{fig3} is
the resulting magnetization curves $M(H)$ of Ba$_3$IrTi$_2$O$_9$ and Ba$_3$RuTi$_2$O$_9$ measured at $T=1.5$~K up to $\mu_0H=55$~T.
With increasing field, $M(H)$ features a steep nonlinear increase at low fields, followed by a quasilinear increase at high fields.
The linear-field dependence of $M(H)$ at higher fields is typical for an antiferromagnetic system  while the upward convex $M(H)$
at low fields is ascribed to weakly correlated orphan spins. This dichotomic characteristic allows analyzing $M(H)$ in terms of a sum of the defect and intrinsic contributions. The orphan spin contribution is not simply described by a Brillouin function $B_J$ since the orphan spins interact weakly to each other in the background of strongly correlated spins. Rather, a modified Brillouin function gives a better description of the data with $M_{\mathrm{imp}}(J,H)=n_\textrm{imp}g\mu_B J B_J[gJH\mu_B/k_B(T-\theta_\textrm{imp})]$,
where $n_\textrm{imp}$ is the number of orphan spins in a unit cell and $\theta_\textrm{imp}$ is related to the correlation temperature of defect spins. $g=2$ is taken from a typical value of the $g$ factor in MO$_6$ (M=Ir$^{4+}$ and Ru$^{4+}$) octahedral environments~\cite{AB}.

For the case of Ba$_3$IrTi$_2$O$_9$, the magnetization curve is fitted to
$M(H)=M_{\mathrm{imp}}(J=1/2,H)+\chi_{\mathrm{intr}}(T)H$.
As evident from Fig.~\ref{fig3}(a), the above equation reproduces nicely
the experimental data. The parameters are estimated to $n_\textrm{imp}=0.053$ and $\theta_\textrm{imp}=1.69$~K. A 5.3 percentage of the defect spins mean that most of the 39~\% Ir$^{4+}$ spins in the Ti(2) site are strongly exchanged coupled to the Ir$^{4+}$ spins
in the magnetic triangular plane through the out-of-plane
interactions, being involved in creating a 2D arrangement of the 1/4-depleted face-sharing tetrahedra. In this sense, the extracted
orphan-spin concentration corresponds to deviation from the exact 1/3 depletion in the Ir$^{4+}$ triangular layer.

Estimating the defect contribution of Ba$_3$RuTi$_2$O$_9$, we check the goodness of fit for various choices of $n_\textrm{imp}$ in a field range of $\mu_0H\in [0,7]$~T. Choosing
$n_\textrm{imp}>0.02$ ends up in a physically irrelevant situation where the theoretical curve overshoots the low-field $M(H)$.
An upper limit of the magnetic impurity content does not exceed 2~\%  impurities. We obtain the best
fit with a sum of 1.7 \% of the orphan spins and a sublinear behavior, $\chi_{\mathrm{intr}}(T)H^{0.719}$.
It is worth noting that the sublinear magnetization has been reported in systems with bond randomness~\cite{Do}. Possibly,  the two exchange interactions $J_1$ and $J_2$ are no longer treated
as identical unlike Ba$_3$IrTi$_2$O$_9$.

\subsection{Low-temperature magnetic specific heat}
%%%%%%%%%%%%%%%%%%%%%%%%%%%%%%%%%%%%%%%%%%%%%%%%%%%%%%%%%%%%%%%%%%%%%%%%%%%%%%%%%%
%%%%%%%%%%%%%%%%%%%%%%%%%%%%%%%%%%%%%%%%%%%%%%%%%%%%%%%%%%%%%%%%%%%%%%%%%%%%%%%%%

\begin{figure}
\label{figure1}
\centering
\includegraphics[width=8cm]{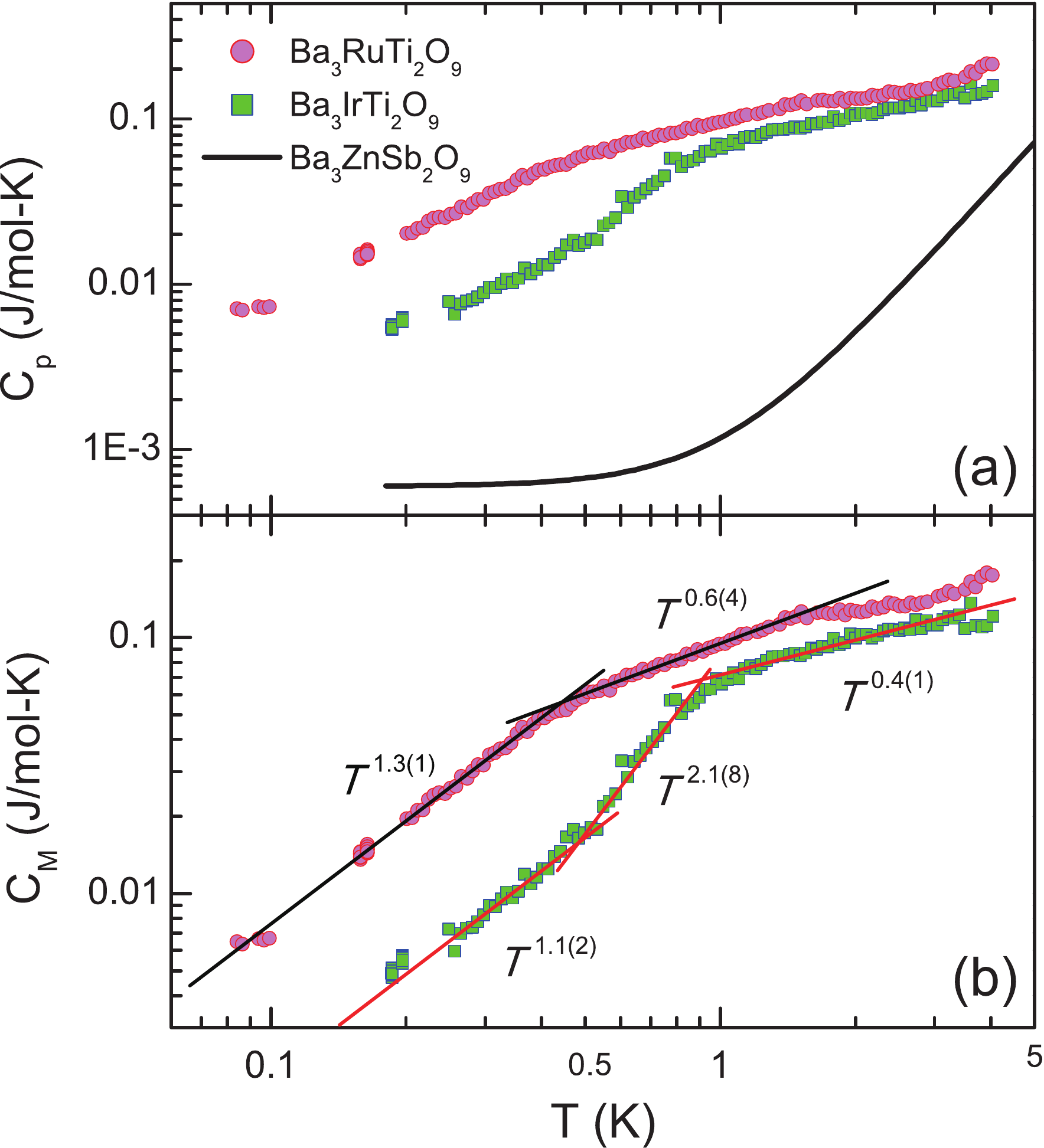}
\caption{(a) Temperature dependence of specific heat of the Ba$_3$MTi$_2$O$_9$
polycrystalline samples plotted on a log-log scale together with the non-magnetic counterpart Ba$_3$ZnSb$_2$O$_9$ (thick solid line).
The specific heat measurements were performed for temperatures below 4 K and at the zero magnetic field. (b)
The magnetic specific heat $C_m$ of Ba$_3$MTi$_2$O$_9$, obtained by subtracting the non-magnetic contribution. The solid lines are the fits of the $C_m$ data to a power-law $\gamma T^{\alpha}$.}
\end{figure}

Ultra-low-temperature specific heat has proven to be a powerful means to elucidate
the nature of low-energy excitations.
Figure~3(a) shows a log-log plot of the specific heat vs temperature for Ba$_3$MTi$_2$O$_9$
measured in the temperature range of $T=0.08-4$~K at $\mu_0H=0$~T.
For both compounds, we found a substantial magnetic contribution but no signature of long-range magnetic order down to 80~mK.
Moreover, the specific heat does not exhibit a Schottky hump, implying a negligible fraction of the orphan spins.
This lends further support for the proposed 1/4 depleted tetrahedral model.
The magnetic specific heat ($C_m$) is obtained by the substraction of the lattice contribution
(thick solid line) from the raw data.

As plotted in Fig.~3(b), the low-$T$ $C_m$ data below 4 K
were fitted to a power-law  $C_m=\gamma T^{\alpha}$ where $\gamma$ is a constant.
$C_m(T)$ of the Ru compound changes the exponent $\alpha=0.6(4)$ to 1.3(1) upon cooling through $T=0.4$~K.
As to  the Ir compound,  $C_m(T)$  is described by a $T$-linear and a $T$-squared behavior for $T< 0.9$~K,
followed by a sublinear decrease with $\alpha=0.41$. Going from the Ru to the Ir sample,
the asymptotic $T$-linear behavior becomes more well-defined while the $T^2$ dependence newly appears.
From analysis of our $C_m$ data we extract the Sommerfeld coefficient $\gamma=36.(8)$~mJ/mol K$^{2}$
for the Ir compound and $\gamma=126.(3)$~mJ/mol K$^{2}$ for the Ru compound.
Noteworthy is that the obtained values of the Sommerfeld coefficient are on the same order as the reported
QSL candidate materials~\cite{Balents}.

Usually, the $T$-linear term together with the non-vanishing value of $\gamma$
is taken as evidence for the presence of low-energy gapless spinon excitations with a pseudo-Fermi surface~\cite{Lee}. In a case of kagome-lattice spin systems,  the $T^2$ dependence is predicted in a theoretical model, in which spinon excitations obey the Dirac spectrum~\cite{Ryu,Ran}. However,
this quadratic behavior has often been reported in other frustrated magnets, for example, the $S=1$ triangular lattices NiGaS$_2$ and
Ba$_3$NiSb$_2$O$_9$ as well as
the hyperkagome iridate Na$_4$Ir$_3$O$_8$~\cite{Cheng,Nakatsuji05,Okamoto}.
As the Ba$_3$MTi$_2$O$_9$ compounds do not realize a perfect kagome lattice,
the observed linear and quadratic $T$ dependence of $C_m$ cannot be unambiguously interpreted in terms of the gapless spinon excitations. Qualitatively, this can be associated with persisting spin dynamics with unconventional low-lying excitations (yet not excluding spinon-like excitations).

\subsection{ZF- and LF- Muon spin resonance}
%%%%%%%%%%%%%%%%%%%%%%%%%%%%%%%%%%%%%%%%%%%%%%%%%%%%%%%%%%%%%%%%%%%%%%%%%%%%%%%%%%
%%%%%%%%%%%%%%%%%%%%%%%%%%%%%%%%%%%%%%%%%%%%%%%%%%%%%%%%%%%%%%%%%%%%%%%%%%%%%%%%%

\begin{figure*}
\label{figure1}
\centering
\includegraphics[width=16cm]{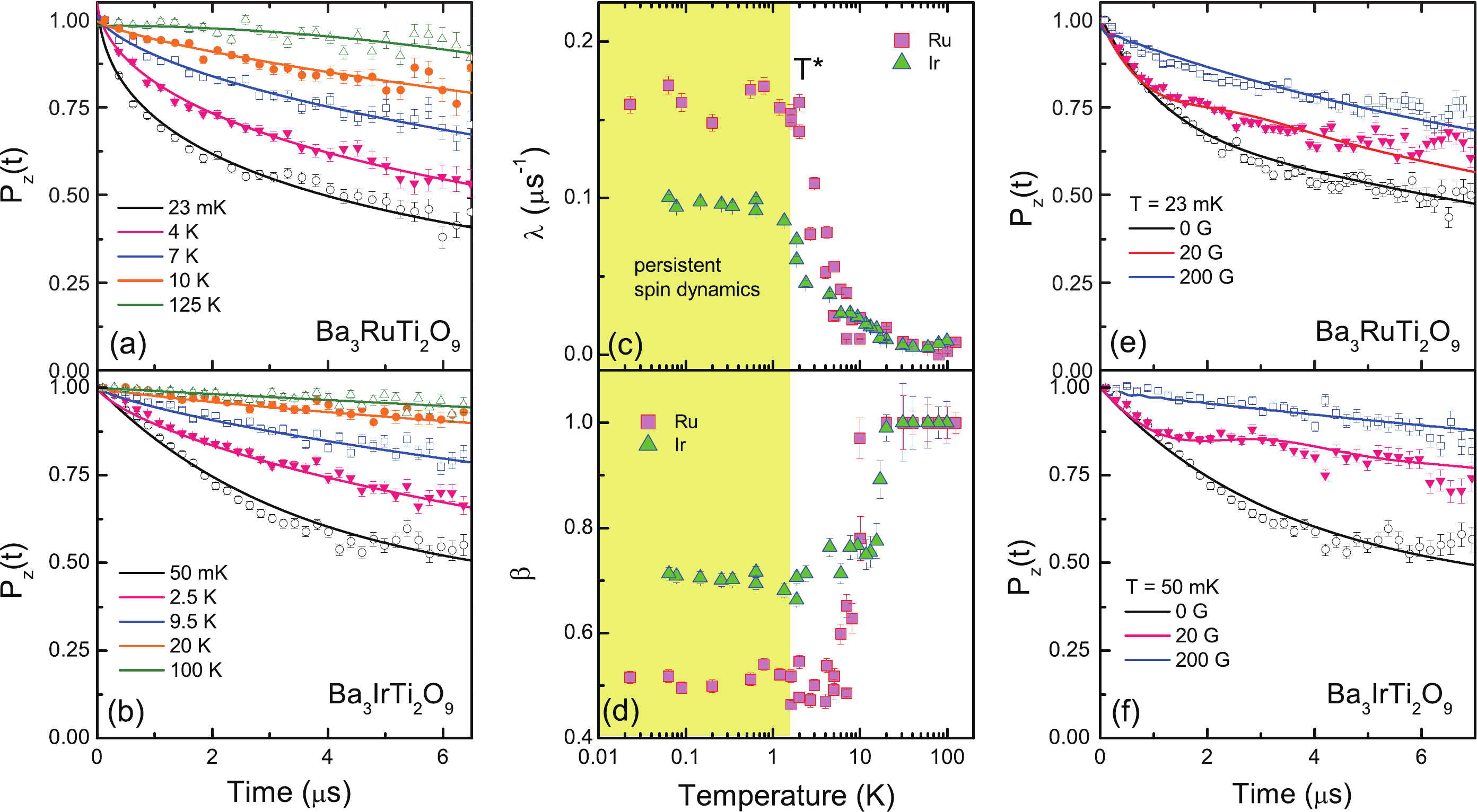}
\caption{ (a),(b) Zero-field muon spin depolarization $P(t)$ of Ba$_3$MTi$_2$O$_9$, measured in the temperature range $T=0.02 - 125$~K. The solid lines are fitted to the stretched exponential function.  (c),(d) Temperature dependence of the muon spin relaxation rate $\lambda_{\mathrm{ZF}}$ and stretching exponent $\beta$ extracted from the zero-field data. (e),(f) Longitudinal-field dependence of the muon spin depolarization measured at $T= 23$ (50)~mK for the Ru (Ir) compound in a magnetic field $H=0- 200$~G. }
\end{figure*}

Muon spin relaxation  ($\mu$SR)  is further employed as a local probe to discriminate between static and dynamic magnetism.
The time-dependent muon spin polarization $P(t)$ of Ba$_3$MTi$_2$O$_9$ in zero external field
is shown in Figs.~4(a) and 4(b) for selected temperatures. Neither spontaneous oscillation nor missing initial
$P(t)$  is observed for all temperatures down to $23 - 50$~mK, confirming the absence of long-range magnetic order.
Rather, the muon spin relaxation is governed by rapidly fluctuating dynamical fields
 on the muon timescale. The zero-field (ZF) $\mu$SR spectra are well described by
the stretched-exponential function $P(t) = f\exp[-(\lambda_{\mathrm{ZF}} t)^\beta]+(1-f)\exp(-\lambda_{\mathrm{bg}} t)$, where \textit{f} is the fraction of the samples, $\lambda_{\mathrm{ZF}}(T)$ is the muon spin relaxation, and $\beta$ is the stretching exponent.
The stretched-exponential function models a magnetic system with a distribution of relaxation rates.
The second simple exponential term accounts for a background contribution to $P(t)$, which stems
from muons implanted in the silver sample holder  or the cryostat. The background relaxation rate $\lambda_{\mathrm{bg}}$ is temperature independent.

The $T$ dependence of the relaxation rate and the stretching exponent extracted from the fitting are plotted
in Figs.~4(c) and 4(d).  At high temperatures above 10~K, the muon spins are weakly relaxing with a simple
exponential relaxation function, indicating that the system is close to its
paramagnetic limit. On cooling below about $T=10$~K,
$\lambda_{\mathrm{ZF}}(T)$ increases strongly, evidencing a slowing down of electronic spin correlations.
As the temperature is further lowered, $\lambda_{\mathrm{ZF}}(T)$ levels off below 1.4 K,  signaling a crossover into a regime characterized by persistent spin dynamics. The so-called persistent slow dynamics has been ubiquitously observed in a range of the QSL candidates, for example, the kagome antiferromagnet ZnCu$_3$(OH)$_6$Cl$_2$~\cite{Mendels}, the triangular lattice YbMgGaO$_4$~\cite{Li}, and the distorted kagome bilayers Ca$_{10}$Cr$_7$O$_{28}$~\cite{Balz}. However, the saturation of the relaxation
should not be regarded as an exclusive property of a QSL because the reported ground states encompass spin freezing, and a weak magnetic order as well~\cite{Yaouanc}. Rather, this behavior is likely linked to unconventional low-energy modes arising from either a QSL
or a weakly symmetry broken state.

Figures 4(e) and 4(f) show the results of longitudinal-field (LF) $\mu$SR measurements where the external field is applied parallel to the initial muon spin direction. $P(t)$ measured at $T= 23$ (50)~mK for the Ru(Ir) compound displays an exponential-type relaxation up to 200~G. When the low-$T$ plateau observed in $\lambda_{\mathrm{ZF}}(T)$ originates from  static magnetism, the local static field
can be estimated to $B_{\mathrm{loc}}=\lambda/\gamma_{\mu}\approx 1.17\,(1.88)$~G for the Ir(Ru) compound, with the muon gyromagnetic ratio  $\gamma_{\mu}/2\pi=135.5~\mathrm{MHz/T}$. In this case, applying a small longitudinal field of about 20~G is sufficient to decouple the muon spins from the samples. However, we find no sign of a full polarization of the LF-$\mu$SR spectra in an external field of 200~G, substantiating a dynamic nature of the magnetism.

We next turn to the evolution of the relaxation shape. As the temperature is lowered, the stretching exponent decreases gradually
from $\beta=1$ to a constant value of about $\beta=0.5$ ($T<1.5$~K) for the Ru compound
and $\beta=0.7$ for the Ir compound. The obtained values of $\beta$ are
larger than the $\beta=1/3$ expected for a canonical spin glass. We note that
the leveling off of the stretching exponent accompanies an order-of-magnitude increase of $\lambda_{\mathrm{ZF}}(T)$
for the Ru compound, which is much smaller than an increase by several orders of magnitude reported in a typical
spin glass~\cite{Uemura}.  This suggests that Ba$_3$RuTi$_2$O$_9$ has a more dynamic nature than spin-glass-like freezing.
Compared to the Ru compound, the Ir compound has the closer value of $\beta$ to 1 and shows the smaller increase of $\lambda_{\mathrm{ZF}}(T)$. Thus, a residual frozen moment is melt down in the Ir compound.

Taken together, the $dc/ac$ magnetic susceptibility of the $4d$ S=1 compound Ba$_3$RuTi$_2$O$_9$ points towards
a spin-glass-like transition. However, the apparent lack of heat capacity anomaly and the persisting spin dynamics seen by $\mu$SR
defy obvious symmetry breaking at temperatures as low as 23~mK. Rather, the magnetic behavior is governed by unconventional low-energy excitations, leading to a power-law dependence of $C_m$. We conclude that although a small fraction of the Ru$^{4+}$ spins are locked into
frozen configurations, the ground state of Ba$_3$RuTi$_2$O$_9$ has a dynamical nature as rationalized in the pyrochlore material Tb$_2$Sn$_2$O$_7$~\cite{Reotier}.

Contrarily, the $5d$ $J_\mathrm{eff}=1/2$ iridate displays no sharp peak and frequency dependence in the $ac$ magnetic susceptibility. In addition, the $T^1$ and $T^2$ dependence of $C_m$ is more well-defined and the muon relaxation rate, while entering a persistent dynamic regime, is less enhanced. These features suggest that the Ir compound has a weaker slow dynamics than the Ru compound. Judging from the fact that both compounds have the same spin topology with a similar level of the site disorders, the small spin number and strong spin-orbit coupling in the iridate are responsible for the stabilization of the purely dynamically fluctuating state in the way that enhanced quantum fluctuations melt the residual frozen moments present in the $4d$ ruthenate.

It is worth commenting that the disorder-free Ba$_3$IrTi$_2$O$_9$ compound is predicted to have a stripy ordered magnetic ground state within the  $J-K-\Gamma$ triangular model~\cite{Catuneanu}.  As the real compound consists of the combined triangle and kagome geometries having
a larger macroscopic degeneracy than the pure triangular lattice, the aforementioned theoretical model is not applicable
to the Ba$_3$IrTi$_2$O$_9$ material. Given the slightly different bond geometry mediating the intra- and inter-triangular exchange interactions (see Supplementary Material B~\cite{SM}), we cannot exclude the possibility that weak randomness in the $J_1$ and $J_2$ interactions promotes the dynamic ground state~\cite{Furukawa}.
{\it ab initio}  calculations of the two $J_1$ and $J_2$ exchange paths are needed to confirm this scenario.
Our work calls for a further theoretical study to clarify the occurrence of an exotic state of matter in the two-dimensional 1/4 depleted tetrahedral lattice.

\section{CONCLUSIONS}
In summary, a combined study of the magnetization, $dc/ac$ magnetic susceptibility, specific heat,
and $\mu$SR on the spin-orbit coupled frustrated antiferromagnets Ba$_3$MTi$_2$O$_9$ reveals that
their magnetic ground state is tuned from the partially frozen state in the ruthenate
to the spin-liquid-like state in the iridate.  This result demonstrates that an intriguing interplay of Kitaev-type interaction
and geometrical structure is conspired to stabilize an exotic state of matter in a triangle-related iridate, deserving future theoretical investigations.

\section*{Acknowledgements}
We thank B.S. Hitti and G.D. Morris for assistance with the $\mu$SR measurements.
This work was supported by the Korea Research Foundation (KRF) grant (Nos. 2009-0093817, 2017-000431, and 2017-0675) funded by the Korea government (MEST). S.J. is supported by the National Research Foundation (NRF) through the Ministry of Science, ICP \& Future Planning (MSIP) (No. 2016K1A4A4A01922028). A portion of this work was performed at the National High Magnetic Field Laboratory, which is supported by the NSF Cooperative Agreement No. DMR-0654118, and by the State of Florida. Pulsed field experiments are supported by EuroMagNET II under the EC Contract 228043.

\section*{References}
%\bibliography{achemso}

\begin{thebibliography}{99}

\bibitem{Balents}
L. Balents, Spin liquids in frustrated magnets, Nature \textbf{464}, 199 (2010).

\bibitem{Savary}
L. Savary  and L. Balents,  Quantum spin liquids: a review,
Rep. Prog. Phys. \textbf{80}, 016502 (2017).

\bibitem{Shimizu}
Y. Shimizu, K. Miyagawa, K. Kanoda, M. Maesato,  and  G. Saito,  Spin liquid state in an organic Mott insulator with a triangular lattice, Phys. Rev. Lett. \textbf{91}, 107001 (2003).

\bibitem{Yamashita08}
S. Yamashita, Y. Nakazawa, M. Oguni, Y. Oshima, H. Nojiri, Y. Shimizu, K. Miyagawa, and K. Kanoda, Thermodynamic properties of a spin-1/2 spin-liquid state in a $\kappa$-type organic salt, Nat. Phys. \textbf{4}, 459 (2008).

\bibitem{Yamashita09}
M. Yamashita, N. Nakata, Y. Kasahara, T. Sasaki, N. Yoneyama, N. Kobayashi, S. Fujimoto, T. Shibauchi, and Y. Matsuda, Thermal-transport measurements in a quantum spin-liquid state of the frustrated triangular magnet $\kappa$-(BEDT-TTF)$_2$Cu$_2$(CN)$_3$, Nat. Phys. \textbf{5}, 44 (2009).

\bibitem{Pratt}
F. L. Pratt, P. J. Baker, S. J. Blundell,	T. Lancaster,	S. Ohira-Kawamura,	C. Baines,	Y. Shimizu,	K. Kanoda,	I. Watanabe, and G. Saito, Magnetic and non-magnetic phases of a quantum spin liquid, Nature \textbf{471}, 612 (2011).

\bibitem{Yamashita10}
M. Yamashita, N. Nakata, Y. Senshu, M. Nagata, H. M. Yamamoto, R. Kato, T. Shibauchi, and Y. Matsuda, Highly mobile gapless excitations in a two-dimensional candidate quantum spin liquid, Science \textbf{328}, 1246 (2010).

\bibitem{SYamashita}
S. Yamashita, T. Yamamoto, Y. Nakazawa, M. Tamura and R. Kato,  Gapless spin liquid of an organic triangular compound evidenced by thermodynamic measurements, Nat. Comm. \textbf{2}, 275 (2011).

\bibitem{Shen}
Y. Shen,	Y.-D. Li,	H. Wo,	Y. Li,	S. Shen,	B. Pan,	Q. Wang,	H. C. Walker,	P. Steffens,	M. Boehm,	Y. Hao,	D. L. Quintero-Castro,	L. W. Harriger,	M. D. Frontzek,	L. Hao,	S. Meng,	Q. Zhang,	G. Chen, and  J. Zhao, Spinon Fermi surface in a triangular lattice quantum spin liquid YbMgGaO$_4$, Nature \textbf{540}, 559 (2016).

\bibitem{Paddison}
J. A. M. Paddison,	M. Daum,	Z. Dun,	G. Ehlers,	Y. Liu,	M. B. Stone, H. Zhou, and M. Mourigal, Continuous excitations of the triangular-lattice quantum spin liquid YbMgGaO$_4$, Nat. Phys. \textbf{13} 117 (2017).


\bibitem{Mendels}
P. Mendels, F. Bert, M. A. de Vries, A. Olariu, A. Harrison, F. Duc, J. C. Trombe, J. S. Lord, A. Amato, and C. Baines, Quantum magnetism in the paratacamite family: towards an ideal kagome lattice, Phys. Rev. Lett. \textbf{98}, 077204 (2007).

\bibitem{Han}
T.-H. Han,	J. S. Helton,	S. Chu,	D. G. Nocera,	J. A. Rodriguez-Rivera, C. Broholm, and Y. S. Lee, Fractionalized excitations in the spin-liquid state of a kagomelattice antiferromagnet, Nature \textbf{492}, 406 (2012).

\bibitem{Fu}
M. Fu, T. Imai, T.-H. Han and Y. S. Lee, Evidence for a gapped spin-liquid ground state in a kagome Heisenberg antiferromagnet, Science \textbf{350}, 655 (2015).

\bibitem{Kitaev}
A. Kitaev,  Anyons in an exactly solved model and beyond, Ann. Phys. \textbf{321}, 2 (2006).

\bibitem{Baskaran}
G. Baskaran, S. Mandal and R. Shankar,  Exact results for spin dynamics and fractionalization in the Kitaev model, Phys. Rev. Lett. \textbf{98}, 247201 (2007).

\bibitem{Knolle}
J. Knolle, D. L. Kovrizhin, J. T. Chalker and R. Moessner,  Dynamics of a two-dimensional quantum spin liquid: signatures of emergent majorana fermions and fluxes, Phys. Rev. Lett. \textbf{112}, 207203 (2013).

\bibitem{Singh}
Y. Singh, S. Manni, J. Reuther, T. Berlijn, R. Thomale, W. Ku, S. Trebst, and P. Gegenwart, Relevance of the Heisenberg-Kitaev model for the honeycomb lattice iridates A$_2$IrO$_3$, Phys. Rev. Lett. \textbf{108}, 127203 (2012).

\bibitem{Takayama}
T. Takayama, A. Kato, R. Dinnebier, J. Nuss, H. Kono, L. S. I. Veiga, G. Fabbris, D. Haskel, and H. Takagi, Hyperhoneycomb iridate $\beta$-Li$_2$IrO$_3$ as a platform for Kitaev magnetism, Phys. Rev. Lett. \textbf{114}, 077202 (2015).

\bibitem{Modic}
K. A. Modic, T. E. Smidt, I. Kimchi, N. P. Breznay, A. Biffin, S. Choi, R. D. Johnson, R. Coldea, P. Watkins-Curry, G. T. McCandless, J. Y. Chan, F. Gandara, Z. Islam, A. Vishwanath, A. Shekhter, R. D. McDonald, and J. G. Analytis, Realization of a three-dimensional spin-anisotropic harmonic honeycomb iridate, Nat. Commun. \textbf{5}, 4203 (2014).

\bibitem{Glamazda}
A. Glamazda, P. Lemmens, S.-H. Do, Y. S. Choi, and K. -Y. Choi, Raman spectroscopic signature of fractionalized excitations in the harmonic-honeycomb iridates $\beta$- and $\gamma$-Li$_2$IrO$_3$, Nat. Commun. \textbf{7}, 12286 (2016).

\bibitem{Plumb}
K. W. Plumb, J. P. Clancy, L. J. Sandilands, V. V. Shankar, Y. F. Hu, K. S. Burch, H.-Y. Kee, and Y.-J. Kim, $\alpha$-RuCl$_3$: a spin-orbit assisted Mott insulator on a honeycomb lattice, Phys. Rev. B \textbf{90}, 041112 (2014).

\bibitem{Sandilands}
L. J. Sandilands, Y. Tian, K. W. Plumb, Y.-J. Kim and K. S. Burch, Scattering continuum and possible  fractionalized excitations in $\alpha$-RuCl$_3$, Phys. Rev. Lett. \textbf{114}, 147201 (2015).

\bibitem{Banerjee}
A. Banerjee, C. A. Bridges,	J.-Q. Yan,	A. A. Aczel, L. Li,	M. B. Stone, G. E. Granroth,	M. D. Lumsden,	Y. Yiu,	J. Knolle,	S. Bhattacharjee,	D. L. Kovrizhin,	 R. Moessner,	D. A. Tennant,	D. G. Mandrus, and S. E. Nagler, Proximate Kitaev quantum spin liquid behavior in a honeycomb magnet, Nat. Mater. \textbf{15}, 733 (2016).

\bibitem{Zhou}
H. D. Zhou, E. S. Choi, G. Li, L. Balicas, C. R. Wiebe, Y. Qiu, J. R. D. Copley and  J. S. Gardner,  Spin Liquid State in the S=1/2 Triangular Lattice Ba$_3$CuSb$_2$O$_9$, Phys. Rev. Lett. \textbf{106}, 147204 (2011).

\bibitem{Nakatsuji}
S. Nakatsuji, K. Kuga, K. Kimura, R. Satake, N. Katayama, E. Nishibori, H. Sawa, R. Ishii, M. Hagiwara, F. Bridges, T. U. Ito, W. Higemoto, Y. Karaki, M. Halim, A. A. Nugroho, J. A. Rodriguez-Rivera, M. A. Green, and C. Broholm, Spin-Orbital Short-Range Order on a Honeycomb-Based Lattice, Science \textbf{336}, 559 (2012).

\bibitem{Corboz}
P. Corboz, M. Lajko, A. M. L\"{a}uchli, K. Penc and F. Mila,  Phys. Rev. X \textbf{2}, 041013 (2012).

\bibitem{Susuki}
T. Susuki, N. Kurita, T. Tanaka, H. Nojiri, A. Matsuo, K. Kindo, and H. Tanaka, Magnetization Process and Collective Excitations in the S=1/2 Triangular-Lattice Heisenberg Antiferromagnet Ba$_3$CoSb$_2$O$_9$, Phys. Rev. Lett. \textbf{110}, 267201 (2013).

\bibitem{Cheng}
J. G. Cheng, G. Li, L. Balicas, J. S. Zhou, J. B. Goodenough, Cenke Xu, and H. D. Zhou, High-Pressure Sequence of Ba$_3$NiSb$_2$O$_9$ Structural Phases: New S=1 Quantum Spin Liquids Based on Ni$^{2+}$, Phys. Rev. Lett. \textbf{107}, 197204 (2011).


\bibitem{Dey}
T. Dey, A. V. Mahajan, P. Khuntia, M. Baenitz, B. Koteswararao, and F. C. Chou, Spin-liquid behavior in $J_{eff}=1/2$ triangular lattice compound Ba$_3$IrTi$_2$O$_9$, Phys. Rev. B \textbf{86}, 140405(R) (2012).

\bibitem{Dey13}
T. Dey and A. V. Mahajan, Frustration induced disordered magnetism in Ba$_3$RuTi$_2$O$_9$, Eur. Phys. J. B \textbf{86}, 247 (2013).


\bibitem{Kimchi}
I. Kimchi and A. Vishwanath, Kitaev-Heisenberg models for iridates on the triangular, hyperkagome, kagome, fcc, and pyrochlore lattices, Phys. Rev. B \textbf{89}, 014414 (2014).

\bibitem{Becker}
M. Becker, M. Hermanns, B. Bauer, M. Garst, and S. Trebst,  Spin-orbit physics of $j=1/2$ Mott insulators on the triangular lattice, Phys. Rev. B \textbf{91}, 155135 (2015).


\bibitem{Catuneanu}
A. Catuneanu, J. G. Rau, H.-S. Kim and H.-Y. Kee, Magnetic orders proximal to the Kitaev limit in frustrated triangular systems:
Application to Ba$_3$IrTi$_2$O$_9$, Phys. Rev. B \textbf{92}, 165108 (2015).


\bibitem{Furukawa}
T. Furukawa, K. Miyagawa, T. Itou, M. Ito, H. Taniguchi, M. Saito, S. Iguchi, T. Sasaki, and K. Kanoda, Quantum Spin Liquid Emerging from Antiferromagnetic Order by Introducing Disorder, Phys. Rev. Lett. \textbf{115}, 077001 (2015).

\bibitem{Wilhelm}
H. Wilhelm, T. Luhmann, T. Rus and F. A.  Steglich, A compensated heat-pulse calorimeter for low temperatures,
Rev. Sci. Instrum. \textbf{75}, 2700 (2004).


\bibitem{Suter}
A. Suter and B. Wojek,  Musrfit: A free platform-independent framework for $\mu$SR data analysis,  Physics Procedia \textbf{30}, 69 (2012).

\bibitem{SM}
See Supplemental Material at [URL will be  inserted by publisher] for an X-ray powder diffraction and ac suscepbility characterization
of the sample as well as for a detailed information on exchange paths.

\bibitem{Radtke}
G. Radtke, C. Maunders, A. Saul, S. Lazar, H. J. Whitfield, J. Etheridge, and G. A. Botton,
Phys. Rev. B \textbf{81}, 085112 (2010).

\bibitem{Kumar}
R. Kumar, D. Sheptyakov, P. Khuntia, K. Rolfs, P. G. Freeman, H. M. R{\o}nnow, T. Dey, M. Baenitz, and A. V. Mahajan, Ba$_3$M$_x$Ti$_{3-x}$O$_9$ (M = Ir, Rh): A family of 5d/4d-based diluted quantum spin liquids, Phys. Rev. B \textbf{94}, 174410 (2016).


\bibitem{AB}
Abragam, A. and Bleaney, B. \textit{Electron Paramagnetic Resonance of Transition Ions} (Oxford U. P., Oxford, England, 1970).

\bibitem{Do}
S.-H. Do, J. van Tol, H. D. Zhou, and K.-Y. Choi, Phys. Rev. B \textbf{90}, 104426 (2014).

\bibitem{Lee}
S.-S. Lee and P. A. Lee,  U(1) gauge theory of the Hubbard model: spin liquid states and possible application to $\kappa$-(BEDT-TTF)$_2$Cu$_2$(CN)$_3$, Phys. Rev. Lett. \textbf{95}, 036403 (2005).

\bibitem{Ran}
Y. Ran, M. Hermele, P. A. Lee,  and X.-G. Wen, Projected-Wave-Function Study of the Spin-1/2 Heisenberg Model on the Kagom\'e Lattice, Phys. Rev. Lett. \textbf{98}, 117205 (2007).

\bibitem{Ryu}
S. Ryu, O. I. Motrunich, J. Alicea, and Fisher,  P. A. Matthew,  Algebraic vortex liquid theory of a quantum antiferromagnet on the kagome lattice, Phys. Rev. B \textbf{75}, 184406 (2007).


\bibitem{Nakatsuji05}
S. Nakatsuji, Y. Nambu, H. Tonomura, O. Sakai, S. Jonas, C. Broholm, H. Tsunetsugu, Y. Qiu, Y. Maeno, Spin Disorder on a Triangular Lattice, Science \textbf{309}, 1697 (2005).

\bibitem{Okamoto}
Y. Okamoto, M. Nohara, H. Aruga-Katori,  and H. Takagi, Spin-Liquid State in the S=1/2 Hyperkagome Antiferromagnet Na$_4$Ir$_3$O$_8$, Phys. Rev. Lett. \textbf{99}, 137207 (2007).


\bibitem{Li}
Y. Li, D. Adroja, P. K. Biswas, P. J. Baker, Q. Zhang, J. Liu, A. A. Tsirlin, P. Gegenwart, and Q. Zhang, Muon Spin Relaxation Evidence for the U(1) Quantum Spin-Liquid Ground State in the Triangular Antiferromagnet YbMgGaO$_4$, Phys. Rev. Lett. \textbf{117}, 097201 (2016).


\bibitem{Balz}
C. Balz,	B. Lake,	J. Reuther,	H. Luetkens,	R. Schonemann, T. Herrmannsdorfer,	Y. Singh,	A. T. M. Nazmul Islam,	E. M. Wheeler,	J. A. Rodriguez-Rivera,	T. Guidi,	G. G. Simeoni,	C. Baines, and H.  Ryll, Physical realization of a quantum spin liquid based on a complex frustration mechanism, Nat. Phys. \textbf{12}, 942 (2016).

\bibitem{Yaouanc}
A. Yaouanc, P. Dalmas de Reotier, A. Bertin, C. Marin, E. Lhotel, A. Amato, and C. Baines, Evidence for unidimensional low-energy excitations as the origin of persistent spin dynamics in geometrically frustrated magnets, Phys. Rev. B \textbf{91}, 104427 (2015).

\bibitem{Uemura}
Y. J. Uemura, A. Keren, K. Kojima, L. P. Le, G. M. Luke, W. D. Wu, Y. Ajiro, T. Asano, Y. Kuriyama, M. Mekata, H. Kikuchi, and K. Kakurai, Spin Fluctuations in Frustrated Kagome Lattice System SrCr$_8$Ga$_4$O$_{19}$ Studied by Muon Spin Relaxation, Phys. Rev. Lett. \textbf{73}, 3306 (1994).


\bibitem{Reotier}
P. Dalmas de Reotier, A. Yaouanc, L. Keller, A. Cervellino, B. Roessli, C. Baines, A. Forget, C. Vaju, P. C. M. Gubbens, A. Amato, and P. J. C. King,
Spin Dynamics and Magnetic Order in Magnetically Frustrated Tb$_2$Sn$_2$O$_7$,
Phys. Rev. Lett. \textbf{96}, 127202  (2006).




\end{thebibliography}

\end{document}